\documentclass[%
	final,
	journal, 
    comsoc,
	letterpaper,
	oneside,
	twocolumn,
	nofonttune,%
]{IEEEtran}%
\PassOptionsToPackage{usenames,dvipsnames,svgnames,x11names,table,prologue}{xcolor}%
\PassOptionsToPackage{hyphens}{url}%
\PassOptionsToPackage{nomessages}{fp}%
\ifCLASSINFOpdf
  \usepackage[pdftex]{graphicx}
\else
  \usepackage[dvips]{graphicx}
\fi
\ifCLASSOPTIONcompsoc
   \usepackage[caption=false,font=normalsize,labelfont=sf,textfont=sf]{subfig}
\else
   \usepackage[caption=false,font=footnotesize]{subfig}
\fi

\usepackage{stfloats}
\usepackage[english]{babel}%
\usepackage[utf8]{inputenc}%
\usepackage[babel,style=english]{csquotes}%
\usepackage{hyphsubst}%
\usepackage[%
	activate={true,%
	nocompatibility},%
	final,%
	tracking=true,%
	kerning=true,%
	spacing=true,%
	factor=1100,%
	stretch=10,%
	shrink=10%
]{microtype}%
\usepackage{setspace}%
\usepackage{xcolor}%
\definecolor{todonotecol}{RGB}{250,0,0}%
\usepackage{xparse}
\usepackage[%
	colorlinks=false,%
	urlcolor=black,%
	linkcolor=black,%
	citecolor=black,%
	filecolor=black,%
	breaklinks,%
	]{hyperref}%
\usepackage{url}%
\usepackage[%
	acronym,%
	nopostdot,%
	seeautonumberlist,%
	shortcuts,%
	section=chapter,%
	toc,%
]{glossaries}%
\glsaddkey*{url}
{}
{\glsentryurl}
{\Glsentryurl}
{\glsurl}
{\Glsurl}
{\GLSurl}
\glsaddkey*{longpltwo}
{}
{\glsentrylongpltwo}
{\Glsentrylongpltwo}
{\glslongpltwo}
{\Glslongpltwo}
{\GLSlongpltwo}
\newglossaryentry{}{%
	name={},%
	plural={},%
	description={},%
}%
\newacronym{it}{IT}{information technology}
\newacronym{ot}{OT}{operational technology}
\newacronym{opcua}{OPC UA}{Open Platform Communications Unified Architecture}
\newacronym{tsn}{TSN}{time-sensitive networking}
\newacronym{scada}{SCADA}{supervisory control and data acquisition}
\newacronym{vpc}{VPC}{virtual process controller}
\newacronym{plc}{PLC}{programmable logic controller}
\newacronym{udp}{UDP}{User Datagram Protocol}
\newacronym{pubsub}{Pub/Sub}{Publish/Subscribe}
\newacronym{vm}{VM}{virtual machine}
\newacronym{pc}{PC}{personal computer}
\newacronym{ptp}{PTP}{Precision Time Protocol}
\newacronym{ntp}{NTP}{Network Time Protocol}
\newacronym{splc}{Soft-PLC}{Software-PLC}
\newacronym{rte}{RTE}{Real-Time Ethernet}
\newacronym{osi}{OSI}{Open Systems Interconnection}
\newacronym{mac}{MAC}{Media Access Control}
\newacronym{e2e}{E2E}{end-to-end}
\newacronym{tcp}{TCP}{Transmission Control Protocol}
\newacronym{hmi}{HMI}{Human-Machine Interface}
\newacronym{kpi}{KPI}{Key Performance Indicator}
\newacronym{mes}{MES}{Manufacturing Execution System}
\newacronym{rami4.0}{RAMI 4.0}{Reference Architectural Model Industrie 4.0}
\newacronym{tacnet4.0}{TACNET 4.0}{Tactile Internet 4.0}
\newacronym{3gpp}{3GPP}{3rd Generation Partnership Project}
\newacronym{5gppp}{5GPPP}{5G Infrastructure Public Private Partnership}
\newacronym{itu}{ITU}{International Telecommunication Union}
\newacronym{ngnm}{NGNM}{Next Generation Mobile Networks}
\newacronym{agv}{AGV}{automated guided vehicle}
\newacronym{v2v}{V2V}{vehicle-to-vehicle}
\newacronym{v2x}{V2X}{vehicle-to-infrastructure}
\newacronym{d2x}{D2X}{device-to-x}
\newacronym{qos}{QoS}{quality of service}
\newacronym{noa}{NOA}{NAMUR Open Architecture}
\newacronym{dcs}{DCS}{distributed control system}
\newacronym{m2m}{M2M}{machine-to-machine}
\newacronym{sdn}{SDN}{software-defined networking}
\newacronym{erp}{ERP}{enterprise resource planning}
\newacronym{ip}{IP}{Internet Protocol}
\newacronym{lte}{LTE}{Long Term Evolution}
\newacronym{5g}{5G}{5th generation wireless communication systems}
\newacronym{harq}{HARQ}{Hybrid Automatic Repeat Request}
\newacronym{rat}{RAT}{radio access technology}
\newacronym{mc}{MC}{Multi-Connectivity}
\newacronym{fbmc}{FBMC}{Filter Bank Multicarrier}
\newacronym{gfdm}{GFDM}{Generalized Frequency Division Multiplexing}
\newacronym{sla}{SLA}{service-level agreement}
\newacronym{5qi}{5QI}{5G QoS Indicator}
\newacronym{bmbf}{BMBF}{German Federal Ministry of Education and Research}
\newacronym{nrt}{NRT}{non real-time}
\newacronym{irt}{IRT}{isochronous real-time}
\newacronym{rt}{RT}{real-time}
\newacronym{ar}{AR}{augmented reality}
\newacronym{wlan}{WLAN}{wireless local area network}
\newacronym{cpu}{CPU}{central processing unit}
\newacronym{ram}{RAM}{random-access memory}
\newacronym{ue}{UE}{user equipment}
\newacronym{ran}{RAN}{radio access network}
\newacronym{bs}{BS}{base station}
\newacronym{cn}{CN}{core network}
\newacronym{epc}{EPC}{evolved packet core}
\newacronym{mec}{MEC}{mobile edge cloud}
\newacronym{rtt}{RTT}{round-trip time}
\newacronym{vpn}{VPN}{virtual private network}
\glsdisablehyper
\usepackage[%
	backend=biber,%
	style=ieee,%
	isbn=false,%
	hyperref=true,%
	maxbibnames=99,%
	sorting=none,%
	natbib=true,%
	language=english,%
	defernumbers=true,%
	]{biblatex}%
\DeclareFieldFormat{sentencecase}{\csname bbx@colon@search\endcsname#1}

\usepackage{nameref}
\usepackage{soul}
\usepackage{flushend}
\usepackage{textcomp}
\usepackage{calc}
\usepackage{xkeyval}
\usepackage{multirow}
\usepackage{tabulary}
\usepackage{makecell}
\usepackage{verbatim}%
 \usepackage{tikz}
\newcommand{\nl}{\par\noindent} 
\newcommand{\mytilde}{{\raise.17ex\hbox{$\scriptstyle\mathtt{\sim}$}}}

\newlength\textheighttemp%
\newlength\textwidthtemp%
\newlength\textheightstd%
\newlength\textwidthstd%
\newlength\textheightold%
\newlength\textwidthold%
\newlength\tempheight%
\newlength\tempwidth%
\SetProtrusion{encoding={*},family={bch},series={*},size={6,7}}
              {1={ ,750},2={ ,500},3={ ,500},4={ ,500},5={ ,500},
               6={ ,500},7={ ,600},8={ ,500},9={ ,500},0={ ,500}}
\SetExtraKerning[unit=space]
    {encoding={*}, family={bch}, series={*}, size={footnotesize,small,normalsize}}
    {\textendash={400,400}, 
     "28={ ,150}, 
     "29={150, }, 
     \textquotedblleft={ ,150}, 
     \textquotedblright={150, }} 
\SetTracking{encoding={*}, shape=sc}{40}
\makeatletter
\let\blx@rerun@biber\relax
\makeatother
				\newcommand{\disablewr}[1]{#1}%
				\newcommand{\newcommanddisw}[3]{\newcommand{#1}[1]{\disablewr{\textcolor{#2}{#3}}}}%
\renewcommand{\disablewr}[1]{}%
\definecolor{todocol}{named}{red}
\newcommanddisw{\todo}{todocol}{ToDo: #1}%
\definecolor{migucol}{named}{purple}%
\newcommanddisw{\migucom}{migucol}{{@}comment: #1}%
\newcommanddisw{\miguhigh}{migucol}{#1}%

\definecolor{orcol}{named}{blue}%
\newcommanddisw{\orcom}{orcol}{{@}comment: #1}%
\newcommanddisw{\orhigh}{orcol}{#1}%

\definecolor{nisccol}{named}{teal}%
\newcommanddisw{\nisccom}{nisccol}{{@}comment:~\textit{#1}}%
\newcommanddisw{\nischigh}{nisccol}{#1}%

\newcommand{\hide}[1]{}

\begin{document}%
%

\title{%
5G as Enabler for Industrie 4.0 Use Cases: Challenges and Concepts %
\thanks{This research was supported by the German Federal Ministry of Education and Research (BMBF) under grant number KIS15GTI007. The responsibility for this publication lies with the authors. This is a preprint of a work accepted but not yet published at the 23rd IEEE International Conference on Emerging Technologies and Factory Automation (ETFA). Please cite as: M.~Gundall, J.~Scneider, H.~D.~Schotten, M.~Aleksy et al.: “5G as Enabler for Industrie 4.0 Use Cases: Challenges and Concepts”. In: 23rd IEEE International Conference on Emerging Technologies and Factory Automation (ETFA), IEEE, 2018.}
}%
%
%
\author{%
\IEEEauthorblockN{%
    Michael Gundall\IEEEauthorrefmark{1}, %
    Jörg Schneider\IEEEauthorrefmark{1}, %
    Hans D. Schotten\IEEEauthorrefmark{1}, %
    Markus Aleksy\IEEEauthorrefmark{2}, %
    Dirk Schulz\IEEEauthorrefmark{2}, %
    Norman Franchi\IEEEauthorrefmark{3}, %
    	\\%
    Nick Schwarzenberg\IEEEauthorrefmark{3}, %
    Christian Markwart\IEEEauthorrefmark{4}, %
    Rüdiger Halfmann\IEEEauthorrefmark{4}, %
    Peter Rost\IEEEauthorrefmark{4}, %
    Dirk Wübben\IEEEauthorrefmark{5}, %
    Arne Neumann\IEEEauthorrefmark{6}, %
       \\%
  	Monique Düngen\IEEEauthorrefmark{7}, %
    Thomas Neugebauer\IEEEauthorrefmark{8}, %
    Rolf Blunk\IEEEauthorrefmark{9}, %
    Mehmet Kus \IEEEauthorrefmark{9}, %
    Jan Grießbach\IEEEauthorrefmark{10}%
    \\%
}%
\IEEEauthorblockA{%
    \IEEEauthorrefmark{1}German Research Center for Artificial Intelligence GmbH (DFKI), Kaiserslautern, Germany%
	\\%
  \IEEEauthorrefmark{2}ABB Corporate Research Germany, Ladenburg, Germany%
    \\%
    \IEEEauthorrefmark{3}Vodafone Chair Mobile Communications, Technische Universität Dresden, Dresden, Germany%
    \\%
  \IEEEauthorrefmark{4}Nokia Bell Labs, Munich, Germany%
	\\%
   \IEEEauthorrefmark{5}University of Bremen, Bremen, Germany%
	\\%
    \IEEEauthorrefmark{6}inIT - Institute Industrial IT, Ostwestfalen-Lippe University of Applied Sciences, Lemgo, Germany%
	\\%
    \IEEEauthorrefmark{7}Robert Bosch GmbH, Hildesheim, Germany%
	\\%
    \IEEEauthorrefmark{8}Götting KG, Lehrte, Germany%
	\\%
        \IEEEauthorrefmark{9}OTARIS Interactive Services GmbH, Bremen, Germany%
	\\%
        \IEEEauthorrefmark{10}NXP Semiconductors Germany GmbH, Hamburg, Germany%
	\\%
    Email: %
        \{michael.gundall, joerg.schneider, Hans\_Dieter.Schotten%
        \}@dfki.de, %
        \{markus.aleksy, dirk.schulz\}@de.abb.com, %
       \\%
        \{norman.franchi, nick.schwarzenberg\}@tu-dresden.de, %
       \\%
       \{Christian.markwart, Ruediger.halfmann, Peter.m.rost\}@nokia-bell-labs.com, %
       \\%
       wuebben@ant.uni-bremen.de, %
       arne.neumann@hs-owl.de, %
       Monique.Duengen@de.bosch.com, %
       \\%
       neugebauer@goetting.de, %
       rolf.blunk@otaris.de, %
       mehmet.kus@otaris.de, %
       jan.griessbach@nxp.com  
}%
}%


%

%
%
%
%
%
%
%
%
\maketitle
%
%
%
%
%
\begin{abstract}%
The increasing demand for highly customized products, as well as flexible production lines, can be seen as trigger for the “fourth industrial revolution”, referred to as “Industrie~4.0”. Current systems usually rely on wire-line technologies to connect sensors and actuators. To enable a higher flexibility such as moving robots or drones, these connections need to be replaced by wireless technologies in the future. Furthermore, this facilitates the renewal of brownfield deployments to address Industrie 4.0 requirements. 

This paper proposes representative use cases, which have been examined in the German \gls{tacnet4.0} research project. In order to analyze these use cases, this paper identifies the main challenges and requirements of communication networks in Industrie 4.0 and discusses the applicability of \gls{5g}.


\end{abstract}%
\begin{IEEEkeywords}%
TACNET 4.0, Industrie 4.0, 5G, industrial communication, KPI
\end{IEEEkeywords}%
%
%
%
%
%
\IEEEpeerreviewmaketitle
%
%
%
%
%
%
%
%
\section{Introduction}%
\begin{figure}[!t]
\centerline{\includegraphics[width=\columnwidth, trim = 3 0 0 0, clip]{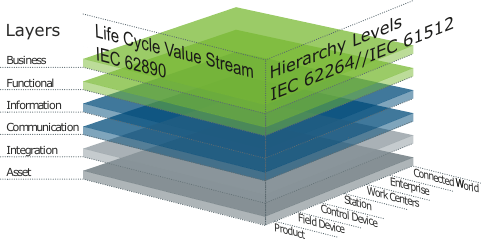}}
	\caption{Visualization of the \gls{rami4.0} \cite{rami40}}
\label{fig:rami}
\end{figure}

\label{sec:Introduction}
Digitalization has become an important topic in industrial environments. Industrie 4.0 describes the “fourth industrial revolution” which enables the customization of products, the flexibility of production lines, and the efficiency of factories \cite{b1}. For this, new automation, information processing, and communication technologies are needed as indicated in the corresponding layers of the \gls{rami4.0} that is shown in Figure \ref{fig:rami}. 
A key objective is to provide a communication layer that supports the seamless access to information related to any type of product or production asset – from sensors to data-analytics services – which is stored in the so-called Industrie 4.0 administration shell \cite{proposaladministrionshell}. The Industrie 4.0 administration shell is the digital representation of all data and functions of a particular product or production asset within an organization, accessible over the network in a uniform, standardized manner. It enables the discovery, negotiation, supervision and use of the production assets \cite{administrationshell}. 

Beside new (“greenfield”) deployments, also the renewal of existing (“brownfield”) facilities requires concepts to add new automation technologies. Typical applications are remote diagnostics and maintenance, logistics, process automation, and remote control, but also novel use cases are foreseen such as the usage of drones, digital twins, mobile assistance systems for human-machine-interaction, or mobile robots, which require new solutions for wireless connectivity.

In order to facilitate the introduction of wireless communication systems, which meet the stringent requirements of industrial deployments, the \gls{bmbf} initiated the collaborative project \gls{tacnet4.0} \cite{tacnet40}. The goal of \gls{tacnet4.0} is the development of a unified industrial \gls{5g} communication system, which is integrated in industrial communication networks. For this purpose, 5G concepts with innovative industry-specific approaches, cross-network adaptation mechanisms and open interfaces between industrial and mobile radio systems are developed. 5G technologies offer concepts that will enable the \gls{tacnet4.0} project to develop efficient solutions for the manufacturing industry. This includes network slicing, flexible frequency spectrum usage, edge cloud concepts, \gls{d2x} communication, private networks, and many more. 
To define and formalize the requirements of industrial use cases, \gls{tacnet4.0} has examined five representative use-cases (see Table \ref{tab:Assignment of the representative use cases to use case groups}) which have partly been defined in the \gls{3gpp} Study on Communication for Automation in Vertical Domains \cite{3gpptr22804}.

In this paper, we present the first project results and describe further steps. Section \ref{sec:Industry 4.0 Use Cases} explains the above-mentioned use cases in detail, motivates their selection, and extracts all relevant \glspl{kpi} with corresponding values. Section \ref{sec:Vertical Communication in Industry 4.0} describes the current deployments as well as challenges and concepts for vertical communication in Industrie 4.0. Based on these results, Section \ref{sec:Current Status and Challenges in Research} specifies the impact on the required functionalities as well as applicable technologies and hardware. Furthermore, the relation between used spectrum range and \glspl{kpi} is analyzed. Finally, Section \ref{sec:Conclusion} concludes the paper.

\section{Industry 4.0 Use Cases}%
\label{sec:Industry 4.0 Use Cases}
Various organizations and authors introduced different approaches to categorize \gls{5g} use cases in the past. The most relevant approaches are defined by the \gls{itu} \cite{ituarchitecture}, \gls{ngnm} Alliance \cite{ngnmarchitecture}, and \hide{Haerick and Gupta (as representatives for the} \gls{5gppp} \cite{5gpparchitecture}. 

Although the use case classification of \gls{5gppp} has commonalities to ours, they do not describe the scope of the \gls{tacnet4.0} project in a proper way. The assignment of the representative \gls{tacnet4.0} use cases to dedicated use case groups is depicted in Table \ref{tab:Assignment of the representative use cases to use case groups}. Subsequently, the considered use cases will be presented in detail providing a short description, an overview of identified benefits and opportunities as well as risks and challenges, and a set of requirements, which complements the use case.

\begin{table}[!t]
\caption{Assignment of the representative use cases to use case groups}
\begin{center}
\begin{tabulary}{\columnwidth}{|C|C|}
\hline 
\textbf{Use case title } & \textbf{Use case group} \\
\hline
 Cooperative Transport of Goods & Mobile Robotics \\
 \hline
 Closed Loop Motion Control & Local and Time Critical Control \\
\hline
 Additive Sensing for Process Automation & Monitoring \\
\hline
 Remote Control for Process Automation & Remote Control \\
\hline
 Industrial Campus & Shared Infrastructure and Intra/Inter Enterprise Communication \\
\hline
\end{tabulary}
\label{tab:Assignment of the representative use cases to use case groups}
\end{center}
\end{table}

\subsection{Cooperative Transport of Goods}%
\begin{figure}[!t]
\centerline{\includegraphics[width=\columnwidth]{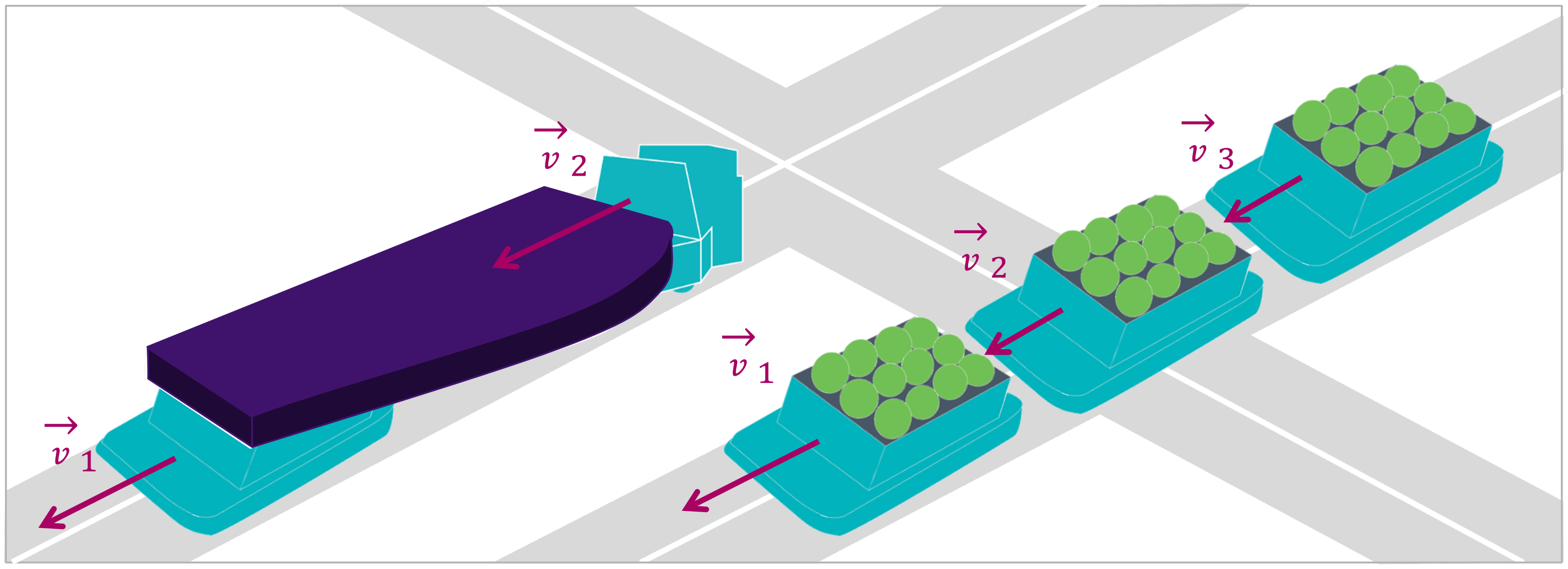}}
	\caption{Cooperative transport of goods (left) and platooning (right)}
\label{fig:platooning}
\end{figure}

\label{subsec:Cooperative Transport of Goods}
This use case describes the cooperative transport of goods and platooning, which means the closed loop control of one vehicle relative to one or more other vehicles. Both scenarios are depicted in Figure \ref{fig:platooning}. 
One task in industrial manufacturing plants is the transport of huge components, e.\,g., wings of wind turbines. Although they are not very heavy, they are too big to be transported by one single driverless \gls{agv} in indoor and outdoor scenarios. Alternatively, a small but very heavy load can be transported by multiple \glspl{agv} as a reconfigurable transport platform, where the \glspl{agv} are combined by virtual drawbars, which is also shown in Figure \ref{fig:platooning}. The relative localization and navigation of the vehicles must be determined through a radio link in a low latency and highly reliable way to avoid unwanted stops or misalignments, which would lead to damages of the transported good or to accidents in a platooning application. An exemplary application for platooning can be found in \cite{daimleragv}, where several driverless trucks equipped with snowplows follow the platoon leader on an airport site.

Today’s applications are realized with radio technologies in accordance to the standard IEEE 802.11p \cite{5514475}. This standard covers the field of \gls{v2v} or \gls{v2x} communication and is state of the art. These radios have cycle times of 100 ms to 200 ms and a radio coverage of several hundreds of meters. Their cycle time is too high for closed loop applications and the coverage is too small for longer platoons. To cover wider indoor and outdoor areas with low latencies and a guaranteed \gls{qos}, it is recommended to use 5G as communication system. With an infrastructure-based communication system, it will also be possible to relocate the steering algorithms to a centralized server and to coordinate several platoons. Furthermore, platoons can also be reconfigured, i.\,e., merging or splitting of platoons. Both the coordination as well as the reconfiguration are necessary to comply with the requirements of Industrie  4.0. 

\subsection{Closed Loop Motion Control}%
\label{subsec:Closed Loop Motion Control}
Motion control is one of the most challenging industrial factory automation use cases and may be open loop or closed loop. In open loop systems, the controller sends a control command and does not receive a feedback while in closed loop systems, feedback is provided and used to initiate dependent actions. In many cases, closed loop control is performed periodically with a defined cycle time interval. The main communication partners involved are a motion control application and one or more sensors and  actuators. 
 \begin{figure}[!t]
\centerline{\includegraphics[width=0.9\columnwidth, trim = 160 310 560 70, clip]{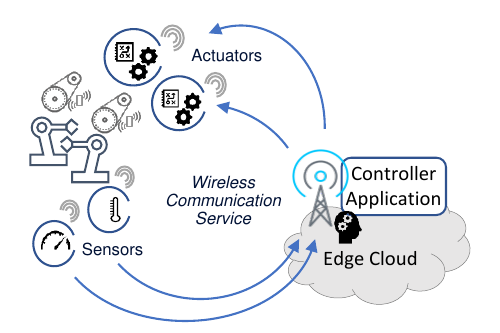}}
	\caption{Closed loop motion control}
\label{fig:Closed Loop Motion Control}
\end{figure}
Figure \ref{fig:Closed Loop Motion Control} shows a typical set-up of closed loop motion control, as it is used for: 

\subsubsection{Machine Tools}%
\label{subsubsec:Machine Tools}
For every machine tool, a master exchanges data via a fieldbus to control many moving components (axes).  For such kind of applications, wireless connectivity may clearly be beneficial. Low fieldbus communication cycle times and a good time synchronization are crucial as command values can change very fast or reactions on actual sensor values must be performed in time.

\subsubsection{Packaging Machines}%
\label{subsubsec:Packaging Machines}
Packaging machines are industrial closed-loop systems, where many moving parts need to be synchronized and coordinated. An example for packaging machines are bottling machines, which clamp up to 15 bottles into a so-called revolver and fill them. The clamping and filling is rotational mounted and must be coordinated at feeding pipe and removal, while it is constantly rotating with a speed of up to 20 m/s.

\subsubsection{Printing Machines}%
\label{subsubsec:Printing Machines}
In an industrial printing system, several moving print heads need to be synchronized with each other and the feed of the paper. The synchronization of the printing line needs to be as exact as possible to avoid a shift of the different colored images and avoid a blurry appearance of the pictures. Thus, the synchronization of the printing cylinders and the material to be printed is most important.

Especially for machine tools and printing machines, the requirements with respect to the timing and reliability are high and thus challenging. Further, the loss of synchronization or messages of only one communication participant can lead to complete production downtime, resulting in high costs. 
\subsection{Additive Sensing for Process Automation}%
\label{subsec:Additive Sensing for Process Automation}
For production automation, there are a number of operational goals with regard to product quality, production uptime, energy and material use, as well as the longevity of production equipment. To optimize toward these goals, insight into process and equipment conditions is needed beyond the information provided by sensors deployed for closed loop control.

By deploying additive sensors for process quantities (temperature, flow, etc.) and equipment conditions (vibrations, leakages, etc.), the sensory resolution in a plant can be significantly increased, including temporary installations to address transient but urgent issues. These sensors typically transmit bursts of data once per hour, where the data depends on the amount of pre-processing within the sensors.

So-called cluster sites hosting multiple (cooperating) process plants may range up to several square kilometers. Already at a sensor density of 1/100 m\textsuperscript{2}, one square kilometer would host a massive amount of 10,000 sensors. To avoid prohibitively high cabling cost (and calendar time in case of temporary deployment), sensor data are transmitted using cellular communication; the proposition is to generally have cellular service available anywhere in the plant, supporting the entire variety of converged applications (see Section \ref{sec:Vertical Communication in Industry 4.0}). Furthermore, the sensors must be energy-autonomous to avoid cabling for power supply as well. Once deployed, they should run for extended periods of time without need for maintenance activities like battery exchange. Given the comparatively low data rates, very large amounts of such sensors can be added without exhausting the resources of the cellular networks (and introducing step costs for network upgrades after all).
 \begin{figure}[!t]
\centerline{\includegraphics[width=\columnwidth, trim = 90 90 380 210, clip]{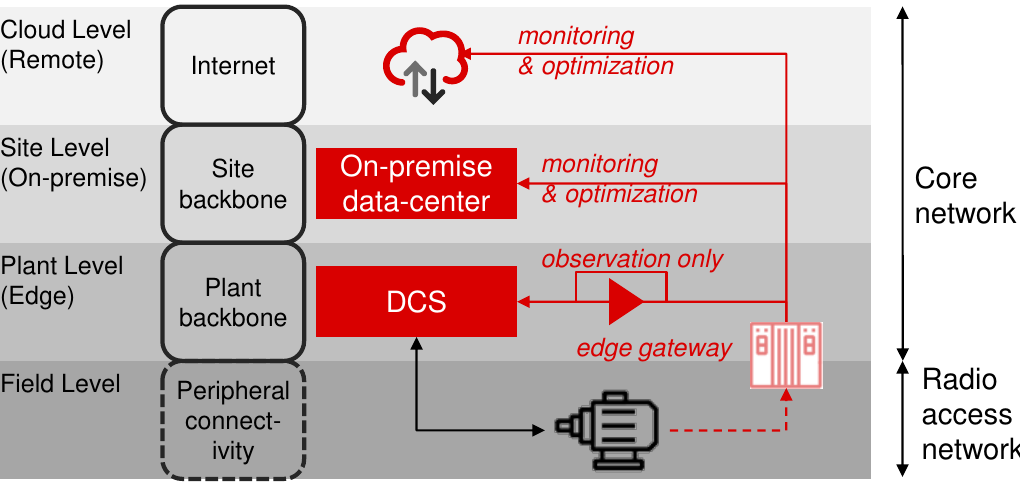}}
	\caption{Using private access and core networks for flexible data routing in monitoring and optimization applications}
\label{fig:Using private access and core networks for flexible data routing in monitoring and optimization applications}
\end{figure}
Depending on customer preference and data privacy concerns, the analytics algorithms may be deployed in an on-premises data-center or in a remote cloud. Figure \ref{fig:Using private access and core networks for flexible data routing in monitoring and optimization applications} illustrates how a private network is used to first capture sensor data over the air into an edge gateway and subsequently route them to any destination where they are needed. In this manner, data privacy is entirely in the hands of the plant owner. Furthermore, the integrity of \gls{dcs} applications is protected by running additive sensing applications in parallel to the \gls{dcs} connectivity as requested by the chemical industry in the so-called \gls{noa} \cite{NamurOpenArchitecture}.

\begin{table*}[!t]
\caption{Requirement specifications for investigated use cases}
\begin{center}
\begin{tabulary}{\textwidth}{|C|C|C|C|C|}
\hline 
\textbf{Requirement} & \textbf{Cooperative transport of goods} & \textbf{Closed Loop Motion Control } &\textbf{Additive Sensing for Proc. Autom.}& \textbf{Remote Control for Proc. Autom.} \\
\hline
 \textbf{Cycle time} & 10 ms & 0.5 ms - 2 ms & 1h - 1 day & 50 ms \\
 \hline
 \textbf{Message size} & 46 Bytes & 20-50 Bytes &  100 Bytes - 10 MBytes &  n/a (video stream)\\
 \hline
 \textbf{Data rate per entity} & 50 kbit/s & 1 Mbit/s - 10 MBit/s & (burst transmission) & 1 Mbit/s - 100 Mbit/s\\
 \hline
 \textbf{Message error rate} & 10\textsuperscript{-7} &10\textsuperscript{-9} - 10\textsuperscript{-8} & 10\textsuperscript{-4} & 10\textsuperscript{-7} \\
 \hline
 \textbf{Latency} & \textless 10 ms & \textless\textless 50\% of cycle time & not in focus & 50 ms \\
 \hline
 \textbf{Distance of entities} & up to several km & up to several 10 m & up to several km & up to several 100 m \\
 \hline
 \textbf{Velocity} & 50 km/h & 2 - 20 m/s & n/a & n/a \\
 \hline
 \textbf{Traffic type} & cyclic, broadcast & cyclic, uni- or multicast & cyclic, uni- or multicast & cyclic, on-demand \\
  \hline
 \textbf{Entity density} & 2 /km\textsuperscript{2} - 30 /km\textsuperscript{2} & 0.1 /m\textsuperscript{2} & Up to 10,000 /km\textsuperscript{2} & 1,000 /km\textsuperscript{2}\\
\hline
\end{tabulary}
\label{tab:Requirement specifications for investigated use cases}
\end{center}
\end{table*}
\subsection{Remote Control for Process Automation}%
\begin{figure}[!t]
\centerline{\includegraphics[width=\columnwidth, trim = 550 80 90 285, clip]{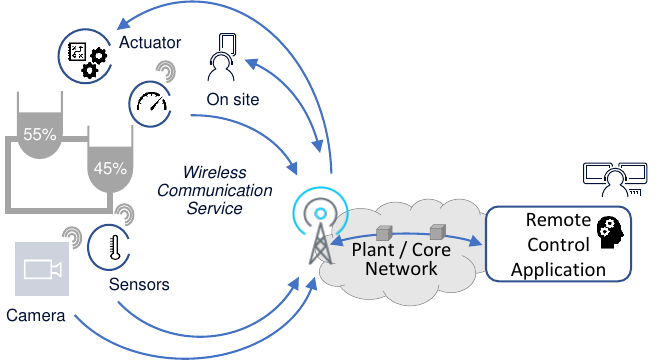}}
	\caption{Remote Control for Process Automation}
\label{fig:Remote Control for Process Automation}
\end{figure}

\label{subsec:Remote Control for Process Automation}
Process automation allows for the automation of (reactive) flows, e.g., refineries and water distribution networks. Process automation is characterized by high requirements on the communications system regarding communication service availability. Systems supporting process automation are usually deployed in geographically limited areas, access to them is usually limited to authorized users, and it will usually be served by private networks. This scenario is depicted in Figure \ref{fig:Remote Control for Process Automation}.

Some of the interactions within a plant are conducted by automated control applications. Therefore, sensor output is requested in a cyclic fashion, and actuator commands are sent via the communication network between a controller and the actuator. Furthermore, there is an emerging need for the control of the plant by on-site personnel. Typically, monitoring and managing of distributed control systems takes place in a dedicated control room.

Staff deployment to the plant itself occurs, for instance, during construction and commissioning of a plant and in the start-up phase of the processes. In this scenario, the local staff taps into the same real-time data as provided to the control room. These remote applications require high data rates because the on-site staff needs to view inaccessible locations with high definition (e.g. emergency valves) and their colleagues in the control room benefit from high-definition footage from body cameras (HD or even 4K). 
Typically, only a few control loops are fully automated and only a handful of control personnel is present on-site, so that the connection density is rather modest.

Table \ref{tab:Requirement specifications for investigated use cases} summarizes typical values for dominant requirements of the aforementioned use cases. This already indicates the large variety of properties to be supported by future communication networks in the context of Industrie 4.0. For the three subtypes of the use case Closed Loop Motion Control, Table \ref{tab:Requirement specifications for investigated use cases} depicts a span of the identified values.

\subsection{Industrial Campus}%
\begin{figure}[!t]
\centerline{\includegraphics[width=\columnwidth, trim = 75 90 100 70, clip]{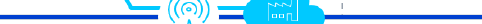}}
	\caption{Using private access and core networks for flexible data routing in monitoring and optimization applications}
\label{fig:Industrial campus with public and private network}
\end{figure}

\label{subsec:Industrial Campus}
Beside the use cases for the specific industrial applications, the use case “Industrial Campus” covers more complex scenarios, which allow for running multiple industrial applications or public and private networks via common network infrastructure. Further, the support of various deployment options for wireless indoor and outdoor communication as well as the handling of multiple companies residing at the same campus is included. This use case is basically defined by operator schemes and procedures rather than by \glspl{kpi}. Therefore, it is not included in Table \ref{tab:Requirement specifications for investigated use cases}. Figure \ref{fig:Industrial campus with public and private network} shows an example how a private and a public mobile service may be provided via a common network infrastructure.
The common infrastructure hosts network resources for the private network (e.g. a remote control application operated by the campus owner) and the public network (e.g. telephony provided by a mobile network operator). This ability is usually referred to as network slicing where multiple customized and isolated logical networks are provided through the same infrastructure \cite{7926920}. Most important requirements for the industrial campus use case are:
\begin{itemize}
\item flexible control and split of network resources to guarantee a specific \gls{qos} for involved communication partners using the common infrastructure,
\item isolation of the private and public networks to provide privacy and basic security, and
\item simple network management and network sharing options to support even complex campus scenarios with sub-networks, multiple operators and applications.
\end{itemize}

\section{Vertical Communication in Industry 4.0}%
\label{sec:Vertical Communication in Industry 4.0}
To identify the challenges for deploying the use cases, resulting by the aforementioned requirements and \glspl{kpi}, we are identifying the gap between the state of the art and Industrie 4.0 scenarios in this section.
\subsection{State of the Art}%
\label{subsec:State of the Art}
Factories and industrial machines are designed for a life-cycle-time of up to 20 years and more. Furthermore, it is most often not profitable to upgrade existing plants with new technologies. Thus, in a practical Industrie 4.0 scenario, there will be a considerable number of long standing facilities.

\begin{figure}[!t]
\centerline{\includegraphics[width=\columnwidth]{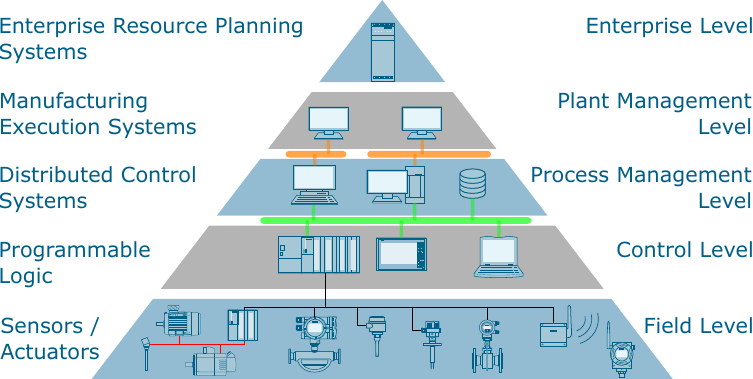}}
	\caption{Automation pyramid}
\label{fig:Automation pyramid}
\end{figure}

Nowadays, the so-called “automation pyramid” dominates the design of industrial communication networks. The automation pyramid shown in Figure \ref{fig:Automation pyramid}, refers to an automation system architecture where automation functions are hierarchically built on top of each other (as reflected in the ISA 95 standard \cite{IEC622641}) and where each layer – from enterprise resource planning to the process equipment – increases in diversity (indicated by width), visually forming a pyramid.
A major challenge is the heterogeneity of industrial communication protocols and interfaces that are located in the lower layers. Especially \glspl{plc} that receive sensor values and control actuators use various communication protocols, which are not necessarily compatible with each other. These so-called fieldbus protocols can differ significantly depending upon use cases, applications and manufacturers. Therefore, applications such as motion control may have very stringent real-time requirements, e.g., regarding the latency of the communication protocol. 
 
For the interface between the \glspl{plc} and the \gls{dcs} located one layer above, larger cycle times may be sufficient. To quantify these aspects, the following real-time classes are usually distinguished \cite{7883994}:
\begin{itemize}
\item real-time class A: $t$\textsubscript{cycle}\textless 100 ms,
\item real-time class B: $t$\textsubscript{cycle}\textless 10 ms, and
\item real-time class C: $t$\textsubscript{cycle}\textless 1 ms.
\end{itemize}

Depending on the required real-time class, proper “Industrial Ethernet”-protocols evolved, e.g., communication protocols addressing real-time class C implement a modified MAC-layer \cite{7883994}, while those addressing real-time class B may be native Ethernet and such addressing real-time Class A may be IP-based. A typical example for an Industrial Ethernet protocol is PROFINET I/O \cite{IEC61784} which contains three traffic classes, \gls{irt}, \gls{rt}, and \gls{nrt}. Therefore, it can address each of the mentioned real-time classes. 

To realize the emerging mobile use cases, there is a necessity for wireless communications. Nowadays, also wireless solutions are used in industrial environments. Typically, these applications do not use mobile radio protocols, but WLAN, Bluetooth, Wireless HART, or ZigBee. Although they cover only a small percentage of applications today, more and more use cases require wireless communication. Comparable to wire-line protocols, each of them has a different advantage regarding required transmission power, coverage, data rate, or resilience. For example, low transmission power is important if the device is battery-powered, or installed in hazardous environments. To achieve high availability, usually the resilience of wireless networks is an important parameter. 

Since most industrial communication protocols are layer 2 protocols (Ethernet-based), at least the according layers of the \gls{osi}-layer model and their interfaces have to be compatible to each other in order to obtain interoperability between constitutive protocols. A first step towards interoperability is the layer 3 (\gls{ip}-based) \gls{opcua} protocol \cite{IEC625411}, developed by the OPC Foundation. It addresses the problem of the heterogeneity of Ethernet-based communication protocols and is a main candidate for the implementation of Industrie 4.0 administration shells. Generally, \gls{ip}-based Industrial Ethernet protocols are suitable for closing the gap between \gls{it} and \gls{ot}, but require new concepts and technologies for optimized routing and message exchange (e.g. \gls{tsn}), especially to achieve real-time class C. The imposed requirements and concepts are introduced in the following section.

\subsection{Challenges and Concepts}%
\label{subsec:Challenges and Concepts}
The convergence of \gls{it} and \gls{ot} in Industrie 4.0 and \gls{5g} leads to a number of integration challenges.
To reduce the effort and cost of integrating and managing a diverse set of technologies, application convergence suggests that a single network technology shall be able to meet the \gls{qos} requirements of any type of application, both for operational (see Section \ref{sec:Industry 4.0 Use Cases}) and \gls{it}  (voice, video, e-mail, file transfer, etc.).

Similarly, to reduce the cost and inflexibility of the operational infrastructure, different types of applications shall concurrently run on top of an over-deployed shared network without violating each other’s security and integrity requirements. To allow the seamless integration of network infrastructures, interfaces used for network (re)configuration and supervision shall be standardized and vendor-independent. 

Toward enabling Industrie 4.0 use cases, a particular challenge is providing information access to Industrie 4.0 administration shells, which act as digital facade for the data and functions of production assets distributed within an organization.
Within an organization, the operational network shall therefore be able to connect endpoints from all levels of an enterprise for vertical integration. Similarly, it shall be able to establish and maintain \gls{e2e} connections across multiple (shared) networks of value-chain partners for horizontal integration; this includes crossing different network operator and security boundaries. 

To provide flexibility for adaptive production, a \gls{sdn} is needed where resources can be reconfigured according to application needs without having to change the physical network.

\begin{figure}[!t]
\centerline{\includegraphics[width=\columnwidth, trim= 205 75 240 110 ,clip]{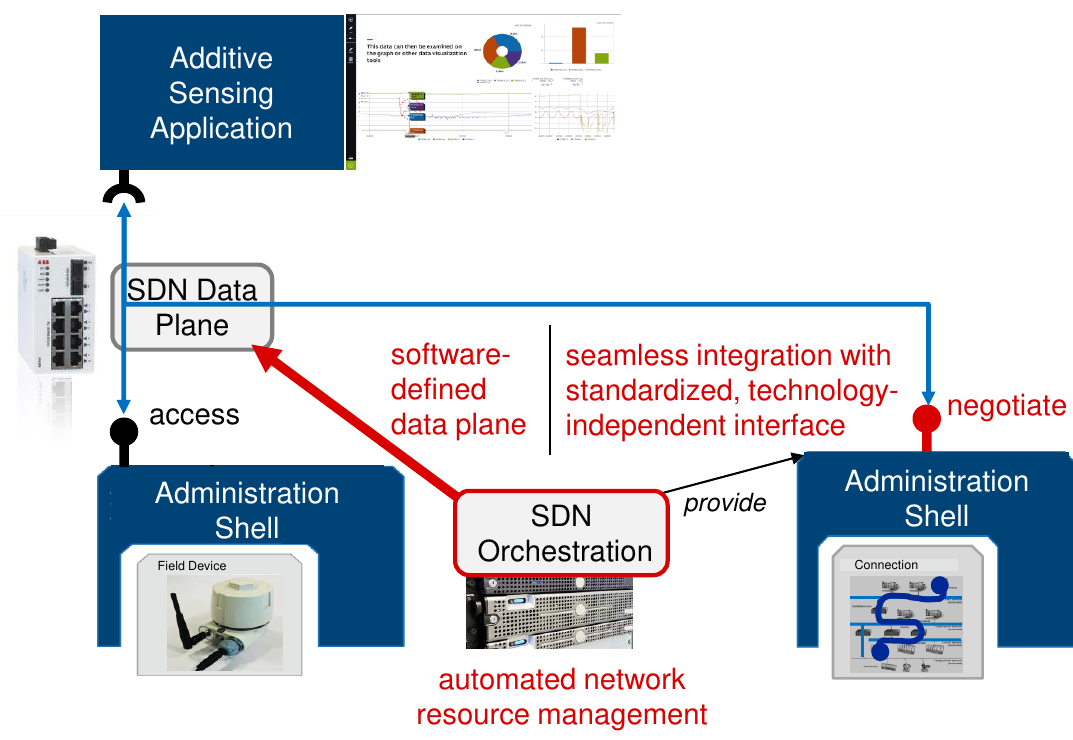}}
	\caption{Additive sensing application negotiating network access to a sensor (refined from \cite{IEC622641})}
\label{fig:Additive sensing application negotiating network access to a sensor}
\end{figure}

Furthermore, to maximize speed and quality of the configuration process, network resources shall self-configure according to the needs of \gls{m2m} applications, i.\,e., without the need for any human input.
 Self-configuring \glspl{sdn} are key enablers for adaptive production  (“plug and produce”). Being mission-critical production assets, Industrie 4.0 would suggest that network connections could be negotiated through corresponding Industrie 4.0 administration shells as indicated in Figure \ref{fig:Additive sensing application negotiating network access to a sensor}. Such a network administration shell would wrap the communication skills of the network using the same principles as for production skills such as measuring, drilling, transporting, etc. In this manner, any type of production application can negotiate its connectivity needs in a technology- and vendor-independent way, only limited by the available resources and \gls{qos} that the specific underlying network and protocol technologies support.
 
The location of the endpoints to be connected in an application context is of particular importance. For connections crossing the boundary of a network segment, e.\,g., when \gls{erp}-level functions connect to production lines or when production lines access information from a material supplier, multiple network resource management functions need to collaborate, sometimes even across company boundaries. The degree of connectedness, where all levels of an enterprise are seamlessly integrated with each other and with partners along the value chain, is what the \gls{rami4.0} calls the Industrie 4.0 Connected World (see figure \ref{fig:rami}).  

\section{Current Status and Challenges in Research}%
\label{sec:Current Status and Challenges in Research}

After describing industrial use cases and their stringent requirements for wireless technologies in Section \ref{sec:Industry 4.0 Use Cases} and showing the state of the art and the target communication networks and technologies with repect to the convergence between \gls{it} and \gls{ot} in Section \ref{sec:Vertical Communication in Industry 4.0}, this section focusses on research approaches concerning the identified gap.

\subsection{New Air Interfaces}%
\label{subsec:New Air Interfaces}

The \gls{tacnet4.0} project analyzes and identifies several \glspl{rat} and modifies, and integrates them. Understanding the properties and applicability of existing technologies is essential in order to develop the radio frequency part of \gls{tacnet4.0} in an efficient way. We are therefore evaluating both, established wireless standards and recent technologies investigated in research projects.

While state of the art cellular equipment already provides sufficient data rates for all of the considered use cases, we identified minimizing the transmission latency as a major challenge. For example, considering \gls{lte} \cite{3GPPTR36881}, typical latencies in downlink and uplink are given by 7.5 ms and 12.5 ms, respectively. Transmission errors are handled by the \gls{harq} scheme by means of repetition, which adds another 8 ms and therefore may exceed latency requirements of industrial applications. While matching \glspl{rat} to the use cases defined in Section \ref{sec:Industry 4.0 Use Cases}, we found that \gls{lte}'s long range and high data rate make it a suitable candidate for “remote control” which has more relaxed latency constraints. Other matches include IEEE 802.11p for “mobile robotics” due to its high mobility and reasonable latency, and ParSec for “local control” thanks to its low cycle times \cite{7760443}.

In addition to making use of existing systems, \gls{tacnet4.0} considers incorporating new concepts for air interfaces of which some shall be presented in the following. The concept of \gls{mc} obviates the need for retransmissions and combines low latency with high reliability by transmitting the same data over multiple independent radio channels. Interestingly, the message error rate is reduced even if the total transmission power is not increased but shared across the available channels \cite{8254232}. Since data is transmitted simultaneously, \gls{mc} does not add latency, but requires additional bandwidth and multiple transmitter-receiver pairs. However, available transmit frequencies are scarce, especially in licensed bands. Considering \gls{lte} or \gls{wlan} transmitters, part of the allocated bandwidth is used as guard band. Advanced waveforms such as \gls{fbmc} and \gls{gfdm} make use of customized filtering to reduce the out-of-band emissions, thereby allowing for smaller guard bands and denser channel allocation. \gls{gfdm} additionally enables flexible adjustment of a transmit symbol’s dimensions in frequency and time. For example, by making a symbol’s subcarriers wider and overlapping, we can improve performance under adverse radio channel conditions as found in factory halls. However, the advantages of \gls{fbmc} and \gls{gfdm} come at the cost of cross-talk between subcarriers and require receivers with advanced equalization algorithms to combat inter-carrier interference \cite{7876875}.

\subsection{New Network Architectures}%
\label{subsec:New Network Architectures}

While most of the \glspl{kpi} of the selected use cases may be fulfilled by the above mentioned modified air interfaces, new concepts for the mobile radio architecture in industrial applications have to be applied. Therefore, we have to integrate \gls{5g}. Here, the major evolution is the so-called operator schemes, where a distinction is made between public and private mobile networks. In general, both types of networks and how they are operated are different and handled independently. Public mobile networks are in general available to all customers being subscribed to the respective mobile network operator. However, it is already possible to operate specific public mobile networks with closed subscriber groups such as for enterprises. In this case, still the same infrastructure of a public mobile operator is used but specific security features or \glspl{sla} may be applied. By contrast, private networks apply to a closed subscriber group and use (partly) private mobile network infrastructure, i.e., a dedicated logical network is implemented for the tenant such as through network slices.

In an industrial network, both public and private networks are applicable, dependent on the specific application. To operate both private and public networks, adequate spectrum resources and highly agile network management with capabilities for self-optimization to meet the needs of all industrial use cases dynamically, are required. Furthermore, it does not matter whether the private \gls{5g} network is planned, built up and operated by the user himself or a service provider commissioned by him. Thus, with \gls{5g} there will also be opportunities and business models based on collaboration between several public, private or even virtual network operators, in order to be able to meet the diverse requirements of heterogeneous industrial communication networks. In order to achieve private and virtualized private networks, the \gls{5g} network slicing technologies will play an important role. 

As a first result, we can say that there is not just one communication technology that can be used to fulfill all the mentioned requirements of the use cases at reasonable costs. Especially in brownfield facilities, several wire-line and wireless technologies will complement each other to meet the \glspl{kpi} \cite{7883994}. On the other hand, the requirements are specified from an end user perspective and therefore apply to the complete \gls{e2e} communication path through the heterogeneous network between the interacting distributed user applications. As a consequence, the function of the technology-specific network segments have to be aligned in order to efficiently provide the required network quality. This is the task of network control, management, and orchestration. For the heterogeneous network, the utilization of isolated and technology-specific or even vendor-specific control and management applications is not sufficient. Also the control and management applications have to be aligned. They need interfaces at different layers of the control and management architecture in order to measure and to influence the network within the time frame required by the use cases named in Section \ref{sec:Industry 4.0 Use Cases}. 

\subsection{Introduction of the TACNET 4.0 Controller}%
\label{subsec:Introduction of the TACNET 4.0 Controller}
In context of the \gls{tacnet4.0} project, a \gls{tacnet4.0} controller is investigated, which coordinates the distributed management entities and fills gaps in the overall control and management architecture of the heterogeneous network. The concepts of \gls{sdn} and network slicing will be considered in the design process of the \gls{tacnet4.0} controller, since these approaches allow the flexibility of control and management procedures according to the Industrie 4.0 concept. The building blocks of its architecture comprise a transformation function which maps user-facing requirement descriptions to technology-facing requirement descriptions and a monitoring function which provides a harmonized representation of network and flow conditions. 

As already mentioned before, interoperability and a seamless protocol integration on user and data plane will be a key technology for \gls{5g} in Industrie 4.0. Mobile networks such as \gls{3gpp} \gls{lte} are highly flexible and adaptable to integrate in any \gls{ip}-based network. With \gls{3gpp} \gls{5g} phase 1 and 2, the mobile network will also provide the means to integrate with layer 2 networks as they are usually deployed in the industrial domain. As a part of this integration, a common understanding and interpretation of \gls{qos} models and guarantees have to be applied. For instance, the definition of stream requirements as defined in \gls{tsn} must be matched with the \gls{5qi} parameters used in \gls{3gpp} \gls{5g}. Finally, these \gls{qos} guarantees must be provided across domains.

Another key technology is mobile edge computing where local network infrastructure can be exploited for both mobile network functions as well as processing on application layer. An example are local control loops requiring low latency communication links. In this case, the user plane processing of the mobile network as well as the control algorithm may be co-located on a common edge-cloud infrastructure.

\subsection{Security Challenges}%
\label{subsec:Security Challenges}

As we see, the resulting overall architecture has a bulk of different facets. Therefore it is not suitable to use standard security concepts. Especially the new concepts that emerge with \gls{5g} in Industrie 4.0 environments drive a next generation of requirements in security \& privacy management and operations:
\begin{itemize}
\item security management and operations must be highly flexible and become adapted and automated in near real-time 
\item security operations must be transformed to be predictive and automated using multi-dimensional analytics, threat intelligence and digital assistance.
\end{itemize}

Since vulnerabilities can lead to unforeseen expenses, a careful analysis of the architecture and its according implementation is the key.

The concept of “Security by Design” addresses this issue by analyzing the use cases and proposed architectures at a very early stage to identify possible threats and attack scenarios. This analysis is done in collaborative workshops with the use case owners and security experts and is refined whenever the architecture or the use case changes, to reflect all possible threats. Each threat is rated in accordance to its damage potential, exploitability, etc. For each risk, a feasible way to mitigate or avoid it, is identified. New threats that arise from the mitigation must be considered in the next iteration of the analysis. All results and architectural changes are documented to make security concepts transparent and security related adaptations traceable.

\subsection{Summary}%
\label{subsec:Summary}
\gls{tacnet4.0} aims to improved \glspl{rat} in order to allow industrial use cases.
The development of a domain specific architecture, operator schemes and management and control methods for seamless integration of 5G and a reliable \gls{e2e} performance will accompany this work. Therefore, another task is the investigation of an overall \gls{tacnet4.0} controller, that controls the individual management entities. In addition, \gls{tacnet4.0} follows this “Security by Design” concept by employing the STRIDE and DREAD threat modeling techniques.

\section{Conclusion}%
\label{sec:Conclusion}
This paper detailed challenges, concepts, use cases, and requirements that are imposed by Industrie 4.0 on \gls{5g} communication networks, which would augment or replace existing networking technologies and enable new use cases. While Industrie 4.0 is the key for flexible, efficient, and adaptable industrial automation, \gls{5g} will provide the tools to realize the Industrie 4.0 vision and expand the use case landscape significantly. In the next step, the ongoing \gls{tacnet4.0} project investigates the system architecture as well as the required interfaces, to provide an integration and migration path towards \gls{5g}-enabled Industrie 4.0 use cases. Individualized security tasks accompany the whole design process.




\printbibliography%
\nl

%
%
\end{document}